\documentclass[12pt]{article}
\newcommand{\tauv}{\mbox{\boldmath$\tau$}}
\usepackage{amsfonts}
\usepackage{amssymb}

\def \be {\begin{equation}}
\def \ee   {\end{equation}}
\def \bea {\begin{eqnarray}}
\def \eea   {\end{eqnarray}}

\begin{document}

\title{\vskip -70pt
\begin{flushright}
{\normalsize DAMTP-2005-63} \\
\end{flushright}
\vskip 60pt {\bf Angularly localized Skyrmions} \vskip 20pt}
{\author{ Olga~V.~Manko {\thanks{Email: O.V.Manko@damtp.cam.ac.uk}}
\vspace{.2cm} \\
and
\vspace{.2cm}
\\  Nicholas~S.~Manton{\thanks{Email: N.S.Manton@damtp.cam.ac.uk}}}}
\date{}
\maketitle

\vspace{-0.5cm}

\begin{center}
\textsl{\large{Department of Applied Mathematics and Theoretical Physics
\vspace{.2cm} \\
Wilberforce Road, Cambridge CB3 0WA,
UK}} \\\vspace{0.7cm}

\large{July 13, 2005}
\medskip
\end{center}

\vspace{.1cm}

\begin{abstract} \noindent
{Quantized Skyrmions with baryon numbers $B=1,2$ and $4$ are
considered and angularly localized wavefunctions for them are
found. By combining a few low angular momentum states, one can
construct a quantum state whose spatial density is close to that
of the classical Skyrmion, and has the same symmetries. For the
$B=1$ case we find the best localized wavefunction among linear
combinations of $j=\frac{1}{2}$ and $j=\frac{3}{2}$ angular
momentum states. For $B=2$, we find that the $j=1$ ground state
has toroidal symmetry and a somewhat reduced localization compared
to the classical solution. For $B=4$, where the classical Skyrmion
has cubic symmetry, we construct cubically symmetric quantum
states by combining the $j=0$ ground state with the lowest
rotationally excited $j=4$ state. We use the rational map
approximation to compare the classical and quantum baryon
densities in the $B=2$ and $B=4$ cases.}
\end{abstract}

\newpage

\section{Introduction}
The connection between the quantum and classical descriptions of a
many--body system is an important but rather tricky one. In
nuclei, the existence of a rotational band, a sequence of states
whose energy increases with angular momentum $j$ approximately as
$\frac{\hbar^2}{2I}j(j+1)$, where $I$ represents a moment of
inertia, suggests the existence of a static intrinsic classical
shape to the nucleus which is not spherically symmetric
\cite{Wong}. It is not obvious how this classical shape arises,
and it is hard to predict the shape, but one can partially
reconstruct it from the spectrum.

For a rigid body, the quantum states of various angular momenta
are given by Wigner functions $D_{sm}^j(\alpha,\beta,\gamma)$,
where $\alpha,\beta,\gamma$ are the Euler angles parametrizing the
orientation, $j$ is the total angular momentum and $s,m$ its
components with respect to the body--fixed and space--fixed third
axis. Symmetries of the body constrain the possible $s$--values or
combinations of $s$--values that can occur. A classically oriented
state is a $\delta$--function in the Euler angles. This can be
obtained by taking an infinite linear combination of Wigner
functions. For a body with symmetry, one would take a sum of
$\delta$--functions on a set of orientations related by symmetry
(which are not distinguishable). Even if there is no fundamental
rigid body to start with, one can consider these linear
combinations. Thus, given a rotational band of states, one can
construct a classically oriented state by taking an infinite
linear combination of true quantum states of definite angular
momentum. The properties of this oriented state (e.g. the particle
density) would define the nature of the intrinsic state.

Something like this has been done in certain condensed matter
situations. One may construct a classically oriented state when
all that is rigorously available is quantum states labelled by
angular momentum. Cooper et al.  \cite{Cooer} have studied a model
of rotating states of a Bose condensate trapped in a harmonic well
(whose shape essentially makes the condensate two--dimensional).
By numerically combining precise states over a range of angular
momenta, they have shown that a condensate with vortices can be
obtained. In the rotating frame, these vortices form a static
array, and so are angularly localized despite the rotational
invariance of the problem. The vortices do not really exist in any
of the states of definite angular momentum, but they do in the
combined state, and they can also be physically observed.

This localization depends on the system being large. Ideally, the
moment of inertia should be almost infinite. In that case, the
angular momentum states of different $j$ are almost degenerate,
and the angular localization may be achieved at almost no energy
cost. (Similarly, an object with large mass can be spatially
localized by taking a superposition of momentum states.)

Unfortunately, for nuclei, this is not always a realistic way to
proceed. For large nuclei, like $\rm{Hf^{170}}$, there are many
states in a rotational band, and it is pretty clear that an
intrinsic nuclear shape exists \cite{Wong}. For smaller nuclei,
however, at most a few low--lying states can be identified as
forming a rotational band, and their energy separation is quite
large because the moments of inertia are smaller. Not much is known
about the wavefunctions of the states in the band, so it is hard
to consider linear combinations. Instead it is better to postulate
some intrinsic shape and fit its parameters to data. In this way,
it is found, for example, that the $\rm{Ne^{20}}$ nucleus has a
prolate deformation \cite{tckrw}, but one cannot say it is exactly
a prolate ellipsoid.

An alternative treatment of many--body systems can give angularly
(and spatially) localized states much more easily. This is the
approach based on an effective field theory, for example a
Ginsburg--Landau description of a Bose condensate. Here, classical solutions
of the field equation can naturally exhibit spatial order, for
example an array of vortices. Because of the underlying
symmetries, the classical solution is not unique, but is
parametrized by collective coordinates describing, say, the center
of mass position and angular orientation. Comparison with the
previous discussion suggests that effective field theory can only
be valid for large systems of many particles. To reconstruct
quantum states of definite angular momentum, one may quantize the
collective coordinates; this makes sense if the mass and the
moment of inertia are of finite, but not infinite magnitude.
A critical comparison of exact quantum states and classical solutions
of an effective field theory has been carried out for quantum Hall
ferromagnets by Abolfath et al. \cite{3}.

In this paper, we shall consider the Skyrme model and its
connection with nuclei and their various angular momentum states.
The Skyrme model is an effective field theory of pions,
with a topological quantum number that can be identified with
baryon number \cite{Skyrme}. Skyrme's original idea was that the model
is justified because nuclei can be thought of as made up of a
condensate of many light pions (with a topological winding).
Recently, the justification is based on the idea that each nucleon
is made of $N_c$ quarks, where $SU(N_c)$ is the gauge group of
QCD, so a nucleus of baryon number $B$ is made of a large number,
$N_cB$, of quarks \cite{428}. The Skyrme model has a
semi--rigorous standing if $N_c$ is large, but it is a
controversial matter whether the physical value $N_c=3$ is
sufficiently large.

The classical Skyrme field equation, like that of the Ginsburg--Landau
model, can be solved numerically and much is known about its
minimal energy solutions (especially for pion mass equal to zero)
\cite{45}. Most importantly, the classical shapes of the
solutions, and their symmetries, are known for values of $B$ up to
and beyond 20 (and work is underway to take account of the finite
pion mass, which could have a qualitatively significant effect for
$B\gtrsim 10$). These classical shapes could represent the
intrinsic shapes of nuclei of modest size.

The shapes obtained have no obvious relation to shapes of nuclei
as understood using other models, in particular, models based on
point nucleons. For example, four--nucleon potential models are
used to describe the $\alpha$--particle, and the classical minimum
occurs for a tetrahedral configuration of the nucleons \cite{FBS}.
In the Skyrme model, the solution of minimal energy with $B=4$ has
cubic symmetry.

Our aim in this paper is to bridge the gap between the classical
Skyrmion shapes and the quantum states of nuclei. The traditional
approach has been to quantize the collective coordinates of
Skyrmions, seek the lowest energy states consistent with the
allowed values of the angular momentum, and compare with the
ground state properties of nuclei. This approach has some success
in reproducing the known spins of nuclei, especially for even
baryon number. More recently, a table of allowed angular momenta
for the ground and first excited states of rotationally quantized
Skyrmions has been constructed \cite{Krusch}. However, in these
quantum states of definite angular momentum, the original Skyrmion
shape information is sometimes completely lost.

We cannot consider an infinite linear combination of angular
momentum states, as we expect that large angular momenta will lead
to Skyrmion deformations, or if these are suppressed, then to
infinite energy. Instead, here we shall consider an intermediate
picture. By taking a small combination of low--lying angular
momentum states, we partially reconstruct the shape of the
classical Skyrmion solution. We shall optimize the angular
localization of the Skyrmion within the limited combinations of
states at our disposal. Such a finite combination of states has
finite energy (not necessarily very much higher than the ground
state). If one could investigate theoretically (or experimentally,
although this could be difficult) the same combination of angular
momentum states in another nuclear model, one might see better the
connection with the Skyrme model picture.

For the $B=1$ case we apparently do not have the problem of
orientation because a single Skyrmion has a spherically symmetric density.
However, the Skyrmion still has rotational collective coordinates, and
we will show that a particular combination of $j=\frac{1}{2}$ and
$j=\frac{3}{2}$ states gives the most localized wave function. We also
show that the ground state of the deuteron (the $j=1$ state of the $B=2$
Skyrmion), without an admixture of higher angular momentum states,
retains the toroidal symmetry of the classical solution. Forest et
al. have argued that not only the pure deuteron state, but also
deuteron clusters within larger nuclei, show toroidal structure
\cite{For}.

Finally, we shall show that a combination of $j=0$ and $j=4$
collective states of the $B=4$ cubic Skyrmion gives a state close to
the classically oriented Skyrmion. The same combination of $j=0$
and $j=4$ states in a four--nucleon potential model could be
compared. Now it is inevitable that in the potential model, the
state will have cubic symmetry because of the Wigner functions
involved. In that sense our discussion is purely kinematic.
However, some detailed properties of the state (density, currents)
might show a close similarity with the Skyrme picture.

This paper is restricted to the $B=1,2,4$ cases, and is
organized as follows. Section 2 contains a review of the Skyrme
model (for more details see \cite{Sutcl}). In Section 3 we give an
outline of the rational map approximation for Skyrmions, and its
consequences, and recall the rational maps for the $B=1$, $B=2$
and $B=4$ Skyrmions. In Section 4, we use the $j=\frac{1}{2}$
and $j=\frac{3}{2}$ quantum states of a $B=1$ Skyrmion introduced
in \cite{Adkins}, and find the combination which gives the best localized
wave function. In Sections 5 and 6 we use the rational maps introduced in
Section 3 to find the ``best'' wavefunctions for $B=2$ and $B=4$
Skyrmions, respectively, and calculate the quantum baryon density
in these states.  In Section 7 we briefly discuss the implications
of adding vibrational modes, and summarize our conclusions.

\section{The Skyrme model}

The Skyrme model \cite{Skyrme} is an effective low energy theory
of QCD attempting to treat pions, nucleons and nuclei. The
topological soliton solutions arising from this model can be
interpreted as baryons.

The model is defined by the Lagrangian
\bea
L &=&\int\left
\{\frac{F_{\pi}^2}{16} \textrm{Tr}
(\partial_{\mu}U\partial^{\mu}U^{\dagger}) \right.  \\
\nonumber  &&  \quad + \left.
\frac{1}{32e^2}\textrm{Tr}([\partial_{\mu}UU^{\dagger},
\partial_{\nu}UU^{\dagger}][\partial^{\mu}UU^{\dagger},
\partial^{\nu}UU^{\dagger}]) \right\} d^3x \,,
\eea
where $U(t,\textbf{x})$ is an $SU(2)$--valued scalar field.
$F_{\pi}$ and $e$ are parameters which can be scaled away by using
energy and length units of $F_{\pi}/4e$ and $2/eF_{\pi}$,
respectively. Thus, with the values of $F_{\pi}$ and $e$ as in
\cite{Nappi} our units are related to conventional units via
$$
\frac{F_{\pi}}{4e}=5.58 \, \textrm{MeV}\,, \qquad \qquad
\frac{2}{eF_{\pi}}=0.755 \, \textrm{fm} \,.
$$

Introducing the $su(2)$--valued right current
$R_{\mu}=(\partial_{\mu}U)U^{\dagger}$ and using geometrical
units, the Lagrangian (1) may be rewritten in the concise form
\be
L=\int\left
\{-\frac{1}{2}\textrm{Tr}(R_{\mu}R^{\mu})+\frac{1}{16}\textrm{Tr}
([R_{\mu},R_{\nu}][R^{\mu},R^{\nu}])\right\} \, d^3x \,. \label{1}
\ee
The Euler--Lagrange equation which follows from (\ref{1}) is
the Skyrme equation
 \be
\partial_{\mu}\left(R^{\mu}+\frac{1}{4}[R_{\nu},[R^{\nu},R^{\mu}]]\right)=0 \,.
\label{2}
\ee
Static solutions are the stationary points (either
minima or saddle points) of the energy function
\be
E=\frac{1}{12\pi^2}\int\left\{-\frac{1}{2}\textrm{Tr}(R_iR_i)-
\frac{1}{16}\textrm{Tr}([R_i,R_j][R_i,R_j])\right\}d^3x \,,
\label{3}
\ee
where we have introduced the additional factor $1/12\pi^2$ for convenience.

$U$, at fixed time, is a map from $\mathbb{R}^3$
into $S^3$, the group manifold of $SU(2)$. However, the boundary
condition $U \to 1$ implies a one--point compactification of
space, so that topologically $U$: $S^3 \rightarrow S^3$, where the
domain $S^3$ is identified with
$\mathbb{R}^3\cup\{\infty\}$. As the homotopy group $\pi_{3}(S^3)$
is $\mathbb{Z}$, maps between 3--spheres are indexed by an
integer, which is denoted by $B$. This integer is also the degree
of the map $U$ and has the explicit representation \be
B=-\frac{1}{24\pi^2}\int\varepsilon_{ijk}\textrm{Tr}(R_iR_jR_k) \,
d^3x \,. \ee As $B$ is a topological invariant, it is conserved
under continuous deformations of the field, including time
evolution. This conserved topological charge Skyrme identified
with baryon number.

Static fields of minimal energy, solving the Skyrme equation, are
called multi--Skyrmions (Skyrmions, for short). They have been
constructed numerically for $B$ up to 22 \cite{45}, and the
symmetries of these solutions have been identified. For $B=1$ the
Skyrmion has spherical symmetry, and for $B=2$ toroidal symmetry.
It turns out that Skyrmions have non--trivial discrete symmetries
for $B>2$. Solutions for negative $B$ are obtained by the
transformation $U \to U^{\dagger}$, which preserves the energy.

\section{Rational map ansatz}

In what follows, we will be using the rational map approximation
to Skyrmions \cite{Manton}. Rational maps were first introduced
into the theory of three--dimensional solitons by Jarvis
\cite{Jarvis}, in the context of monopoles, but they prove to be
very useful for Skyrmions as well.

Rational maps are maps from $S^2\rightarrow S^2$, whereas
Skyrmions are maps from $\mathbb{R}^3\rightarrow S^3$. The idea in
\cite{Manton} is to identify the domain $S^2$ of the rational map
with concentric spheres in $\mathbb{R}^3$, and the target of the
rational map $S^2$ with spheres of latitude on $S^3$. A point in
$\mathbb{R}^3$ can be parametrized by $(r,z)$; $r$ denotes radial
distance and the complex variable $z$ specifies the direction. Via
stereographic projection $z$ can be written in terms of usual
polar coordinates $\theta$ and $\phi$ as
$z=\tan(\theta/2)e^{i\phi}$. A rational map may be written as
$R(z)=p(z)/q(z)$, where $p(z)$ and $q(z)$ are polynomials in $z$.
The degree of the rational map, $N$, is the greater of the
algebraic degrees of the polynomials $p$ and $q$. $N$ is also the
topological degree of the map (its homotopy class) as a map from
$S^2 \to S^2$.

The point $z$ on $S^2$ corresponds to the unit vector
\be
{\bf{\hat{n}}}_z=\frac{1}{1+|z|^2}(z+\bar{z},i(\bar{z}-z),1-|z|^2)
\,.
\ee
Similarly, the value of the rational map $R$ is associated
with the unit vector
\be
{\bf\hat{n}}_R=\frac{1}{1+|R|^2}(R+\bar{R},i(\bar{R}-R),1-|R|^2)
\,.
\ee
The ansatz for the Skyrme field, depending on a rational
map $R(z)$ and a radial profile function $f(r)$, is
\be
U(r,z)=\exp(if(r){\bf\hat{n}}_{R(z)} \cdot {\bf{\tauv}}) \,,
\label{4}
\ee
where ${\tauv}=(\tau_1, \tau_2, \tau_3)$
denotes the triplet of Pauli matrices, and $f(r)$ satisfies
$f(0)=\bf\pi$, $f(\infty)=0$.

An $SU(2)$ M\"obius transformation of $z$ corresponds to a
rotation in physical space; an $SU(2)$ M\"obius transformation of
$R$ (i.e. on the target $S^2$) corresponds to an isospin rotation.
Both these transformations of a rational map are symmetries of the
Skyrme model, and both preserve $N$.

It can be verified that the baryon number for the
ansatz (\ref{4}) is given by
\be
B=-\int\frac{f'}{2\pi^2}\left(\frac{\sin{f}}{r}\right)^2
\left(\frac{1+|z|^2}{1+|R|^2}\left|\frac{dR}{dz}\right|\right)^2
\frac{2i \, dzd\bar{z}}{(1+|z|^2)^2}r^2 \, dr \,. \label{6}
\ee
$2i \, dzd\bar{z}/(1+|z|^2)^2$ is equivalent to the usual 2--sphere
area element $\sin{\theta} \, d\theta d\phi$. The angular part
of the integrand,
\be
\left(\frac{1+|z|^2}{1+|R|^2}
\left|\frac{dR}{dz}\right|\right)^2\frac{2i \, dzd\bar{z}}{(1+|z|^2)^2}
\,, \label{6'}
\ee
is precisely the pull--back of the area form
$2i \, dR \, d\bar{R}/(1+|R|^2)^2$ on the target sphere of the
rational map $R$, so its integral is $4\pi$ times the degree $N$
of the map. Therefore (\ref{6}) simplifies to
 \be
B=\frac{-2N}{\pi}\int_0^{\infty}f'\sin^2f \, dr=N \,.
\ee

An attractive feature of the rational map ansatz is that it leads
to a simple energy expression which can be separately minimized
with respect to the rational map $R$ and the profile function $f$
to obtain close approximations to the numerical, exact Skyrmion
solutions, and having the correct symmetries. (The numerical
solutions are in fact best found by starting from the optimal
rational map approximations.)

Indeed, using (\ref{4}) we get the following expression for the
energy (\ref{3}):
\be
E=
\frac{1}{3\pi}\int_0^{\infty}\left(r^2f'^2+2N\sin^2f(f'^2+1)+
{\mathcal I}\frac{\sin^4f}{r^2}\right)dr \,. \label{5}
\ee
Here ${\mathcal I}$ denotes the purely angular integral
\be
{\cal I}=\frac{1}{4\pi}\int\left(\frac{1+|z|^2}{1+|R|^2}
\left|\frac{dR}{dz}\right|\right)^4\frac{2i \, dzd\bar{z}}{(1+|z|^2)^2}
\,,
\ee
which only depends on the rational map $R$. To minimize
$E$, for maps of given degree $N$, one should first minimize
${\mathcal I}$ over all maps of degree $N$. Then the profile
function $f$, minimizing the energy (\ref{5}), may be found by
numerically solving a second order, ordinary differential equation
with $N$ and ${\mathcal I}$ as parameters.

For $B=1$, the rational map is $R(z)=z$, and this reproduces
Skyrme's hedgehog ansatz \cite{Skyrme}, which is exactly satisfied by the $B=1$
Skyrmion. For $B=2$ and $B=4$ the symmetries of the computed
Skyrmions are $D_{\infty h}$ and $O_h$ respectively, and in each
case there is a unique rational map of the desired degree with the
given symmetry, which also minimizes ${\mathcal I}$. They are,
respectively,
\be R(z)=z^2 \,, \qquad
R(z)=\frac{z^4+2\sqrt{3}iz^2+1}{z^4-2\sqrt{3}iz^2+1} \,.
\label{6^}
\ee
In all these cases, we have made a convenient
choice of orientation in presenting the maps.

When quantizing the Skyrme field, we will be interested in the
behavior of the wavefunction with respect to different
orientations of the Skyrmion configurations. Consequently, all the
information we need will be encoded in the angular dependence of
the baryon density (\ref{6'}), which only depends on the rational
map, and the profile function $f$ will not be of much interest for
our purposes.

\section{$B=1$ case}

The $B=1$ Skyrmion is spherically symmetric
and takes the hedgehog form
\be
U_0({\bf{x}})=\exp\{{if(r)\bf\hat{x}\cdot\bf\tauv})\} \,,
\ee
where $f(0)=\pi$ and $f(\infty)=0$.
If $U_0$ is the soliton solution, then $U=AU_0A^{-1}$, where $A$ is
an arbitrary constant $SU(2)$ matrix, is a static solution as well. But, in
order to get solitons which are eigenstates of spin and isospin one
needs to treat $A$ as a collective coordinate. So substitute
$$
U=A(t)U_0A^{-1}(t)
$$
in the Lagrangian (1), where $A(t)$ is an arbitrary
time--dependent $SU(2)$ matrix. The Lagrangian for $A$ is
\cite{Adkins}
\be
L=-M+\lambda\textrm{Tr}(\partial_0A\partial_0A^{-1})\,,
\label{1.1}
\ee
where $M$ is the soliton mass and $\lambda$ is
an inertia constant which may be found numerically.

The $SU(2)$ matrix $A$ can be written as
$A=a_0+i\bf{a}\cdot\bf\tauv$, with $a_0^2+|{\bf{a}}|^2=1$. In terms of
$a_\xi \ (\xi=0,1,2,3)$ the Lagrangian (\ref{1.1}) becomes
\be
L=-M+2\lambda\sum_{\xi=0}^3(\dot a_\xi)^2 \,,
\ee
and after the usual quantization procedure one gets the Hamiltonian
\be
H=M+\frac{1}{8\lambda}\sum_{\xi=0}^3\left(-\frac{\partial^2}{\partial
a_\xi^2}\right) \,.
\ee
Because of the constraint $a_0^2+|{\bf{a}}|^2=1$, the operator
$\sum_{\xi=0}^3\left(\partial^2/\partial a_\xi^2\right)$ is to be
interpreted as the laplacian $\nabla^2$ on the 3--sphere. The
wavefunctions can be expressed as traceless, symmetric,
homogeneous polynomials in the $a_\xi$.

Using the isospin and spin operators
\bea
I_k=\frac{1}{2}i\left(a_0\frac{\partial}{\partial a_k}
-a_k\frac{\partial}{\partial
a_0}-\epsilon_{klm}a_l\frac{\partial}{\partial a_m}\right),
\nonumber \\
J_k=\frac{1}{2}i\left(a_k\frac{\partial}{\partial a_0}
-a_0\frac{\partial}{\partial
a_k}-\epsilon_{klm}a_l\frac{\partial}{\partial a_m}\right),
\eea
Adkins, Nappi and Witten have found the normalized wavefunctions for
neutron, proton and $\Delta$--resonances \cite{Adkins}. The wavefunctions we
require, having opposite $I_3$ and $J_3$ eigenvalues, are
\bea
|n, s_z=\frac{1}{2} \rangle &=&\frac{i}{\pi}(a_0+ia_3) \,, \nonumber \\
|p, s_z=-\frac{1}{2} \rangle &=&-\frac{i}{\pi}(a_0-ia_3) \,,\nonumber \\
|\Delta^{-},
s_z=\frac{3}{2}\rangle &=&\frac{\sqrt2}{\pi}(a_0+ia_3)^3 \,, \nonumber \\
|\Delta^{0}, s_z=\frac{1}{2}\rangle
&=&-\frac{\sqrt2}{\pi}(a_0+ia_3)(1-3(a_1^2+a_2^2)) \,, \nonumber \\
|\Delta^{+}, s_z=-\frac{1}{2}\rangle
&=&-\frac{\sqrt2}{\pi}(a_0-ia_3)(1-3(a_1^2+a_2^2)) \,, \nonumber \\
|\Delta^{++}, s_z=-\frac{3}{2}\rangle
&=&\frac{\sqrt2}{\pi}(a_0-ia_3)^3 \,. \label{NucleonDelta}
\eea
But none of these
"pure" $j=1/2$ and $j=3/2$ states is the best localized
wavefunction. This will instead be given by a superposition of the
above states.

If we could take into account an infinite number of angular
momentum states the most localized wavefunction would be the Dirac
delta function, which may be expressed in the following form
\be
\delta(\mu)=\sum_j{(2j+1)\chi^{j}(\mu)} \,, \quad \quad
j=0,\frac{1}{2},1,\frac{3}{2},\dots \,, \label{delta}
\ee
where
$\chi^{j}(\mu)$ is the character of the representation of
dimension $(2j+1)$, and $a_0 = \cos\mu$. However, this
wavefunction does not respect the Finkelstein--Rubinstein (FR)
constraints \cite{FR}, which in the case of one Skyrmion requires
that the wavefunction is antisymmetric under $A \to -A$, thus
ensuring that the quantized Skyrmion is a fermion. The sum in
(\ref{delta}) must therefore be restricted to half--integer values
of $j$, giving the total $\frac{1}{2}(\delta(\mu) -
\delta(\mu - \pi))$.

For the $SU(2)$ group, the representation matrices are
matrices of Wigner functions, i.e. for each $j$
$$
{\bf D}^j=\left(
\begin{array}{cccc}
D^j_{jj}&\ldots&D^j_{j-j}\\\vdots&\ddots&\vdots\\
D^j_{-jj}&\ldots&D^j_{-j-j}
\end{array}\right) \,.
$$
In what follows we will be interested in the $j=1/2$ and $j=3/2$
cases, for which the Wigner functions take a concise form in terms
of $a_0,\dots,a_3$. The character $\chi^{j}$ is the trace of the above matrix,
consequently let us write down the diagonal elements:
\bea
D^{1/2}_{1/2,1/2}&=&a_0+ia_3 \,, \nonumber \\
D^{1/2}_{-1/2,-1/2}&=&a_0-ia_3 \,, \nonumber \\
D^{3/2}_{3/2,3/2}&=&(a_0+ia_3)^3 \,, \nonumber \\
D^{3/2}_{1/2,1/2}&=&(a_0+ia_3)(1-3(a_1^2+a_2^2)) \,, \nonumber \\
D^{3/2}_{-1/2,-1/2}&=&(a_0-ia_3)(1-3(a_1^2+a_2^2)) \,, \nonumber \\
D^{3/2}_{-3/2,-3/2}&=&(a_0-ia_3)^3 \,.
\eea
If we truncate the sum (\ref{delta}) at $j=3/2$ we get the following
candidate for a well localized (normalized) wavefunction,
\be
\Psi(a_0,a_1,a_2,a_3)=\frac{8}{\pi}\sqrt{\frac{2}{5}}\left(a_0^3-\frac{3}{8}
a_0\right)\,.
\ee
In terms of nucleon and $\Delta$--resonance states this
can be written as
\be
\frac{1}{\sqrt5}\left(|\Delta^{-}\rangle - |\Delta^{0}\rangle -
|\Delta^{+}\rangle + |\Delta^{++}\rangle
- \frac{i}{\sqrt2} (|n \rangle -|p \rangle)\right) \,,
\ee
with spins as in (\ref{NucleonDelta}).
A more general wavefunction of this type is \be
\Psi(a_0,a_1,a_2,a_3)=\frac{\sqrt2}{\pi}
\left(\frac{5}{16}+\kappa+\kappa^2\right)^{-1/2} \left(a_0^3 +
\kappa a_0\right) \,. \ee The maximum magnitude of $\Psi$ at $a_0
= \pm 1$ occurs when $\kappa = -3/8$, confirming that this is the
best localized wavefunction.

Another measure of how well the wavefunction is localized around
$a_0 = \pm 1$ is given by the integral
\be
\frac{2}{\pi^2}\left(\frac{5}{16}+\kappa+\kappa^2\right)
^{-1}\int_0^{\pi}a_0^2 \, |a_0^3+\kappa a_0|^2 \, d\Omega \,.
\label{intcriterion}
\ee
Here $a_0 = \cos\mu$ and $d\Omega=4\pi\sin^2\mu \, d\mu $ is the
measure of integration. After an easy calculation, we find that
this integral is maximal when $\kappa=-1/4$, which is close to the
result we got before. One more wavefunction worth considering is
$$\Psi=\frac{4}{\pi}\sqrt{\frac{2}{5}}a_0^3\,,$$ which is as well
localized as the one with $\kappa=-3/8$ according to criterion
(\ref{intcriterion}), and rather simpler. It is the following combination of
nucleon and $\Delta$ states:
\be
\frac{1}{2\sqrt5}\left(|\Delta^{-}\rangle - |\Delta^{0}\rangle -
|\Delta^{+}\rangle + |\Delta^{++}\rangle - 2\sqrt2 \, i \, (|n
\rangle -|p \rangle)\right)\,.
\ee

These localized states are not physically important for isolated
nucleons; however, they could be useful for modelling nucleons in
interaction. Recent developments have shown that, for example, the
deuteron is not formed from a proton and neutron only, but
probably also contains some amount of $\Delta$--resonances
\cite{delta, delta1}. Therefore, considering a superposition of
states with different angular momenta is definitely physically
meaningful. In \cite{Leese} the deuteron was modelled by a bound
state of Skyrmions in the attractive channel, where the relative
orientation of the Skyrmions was chosen to maximize the attraction
at short range. Such states could be approximated by the combined
$j=1/2$ and $j=3/2$ states we have discussed here. The dependence
of the force between two Skyrmions on their relative orientation is
the classical analogue of the tensor force between nucleons, and
it appears to automatically lead to an admixture of a
$\Delta$--resonance component to each nucleon.

\section{$B=2$ case}

The $B=2$ Skyrmion has $D_{\infty h}$ symmetry and a toroidal
shape \cite{KopS,Man1,Ver}, and is used to describe the deuteron. We take the
symmetry axis to be the third body--fixed axis, and the Skyrmion to be in
its standard orientation if this coincides with the third
Cartesian axis in space. In the rigid body approximation to
quantization, the wavefunction is a function only of the
rotational and isospin collective coordinates. (We ignore the
translational collective coordinates, and set the momentum to
zero.) To make an appropriate quantization we have to impose
FR constraints, which tell us
that the ground state has the quantum numbers $(i,j)=(0,1)$,
where $i$ is the total isospin and $j$ is the total spin. The
wavefunction describing this deuteron state was obtained in \cite{Braaten}.
Since $i=0$, there is no dependence on the isospin collective
coordinates, and the (normalized) state is
\be
\Psi=\sqrt{\frac{3}{8\pi^2}}D^1_{0m}(\alpha,\beta,\gamma) \,.
\ee
Here $\alpha$, $\beta$ and $\gamma$ are the rotational Euler
angles, $D^1_{0m}(\alpha,\beta,\gamma)$ is a Wigner function, and
$m$ is the third component of the space--fixed spin.

In \cite{Leese}, the analysis was extended to include one
vibrational mode of the system, allowing the $B=2$ toroidal
Skyrmion to separate into two $B=1$ Skyrmions. The wavefunction
therefore includes a factor $u(\rho)$, the radial part of the
deuteron wavefunction, which satisfies a radial Schr\"odinger
equation on the interval $[\rho_0, \infty)$, where $\rho_0$ corresponds
to the toroidal configuration. Here, however, we consider only the
rigid body rotational states, and their angular dependence.

Since we are particularly interested in the spatial orientation, we
treat states differing in $m$ as different. The state we are looking for has to
have the same symmetry properties as the classical solution.
Consequently, the desired wavefunction is
\be
\Psi=\sqrt{\frac{3}{8\pi^2}}D^1_{00}(\alpha,\beta,\gamma)
=\sqrt{\frac{3}{8\pi^2}}\cos{\beta} \,,
\label{deut}
\ee
which is axially
symmetric both on the left and on the right (i.e. with respect to
the body--fixed symmetry axis, and the $x_3$-axis in space).
$\Psi$ has its maximum magnitude at $\beta = 0$ and $\beta = \pi$,
corresponding to the Skyrmion in its standard orientation, and
turned up-side down, which is classically indistinguishable after
an isospin rotation.

Now, given the orientational quantum state (\ref{deut})
we may calculate the nucleon density, and
find how quantum effects change the density of the
classical configuration. We find the expression for the baryon
density distribution $\rho_\Psi({\bf{x}})$ in physical space
(which is interpreted as nucleon density) by averaging the
classical baryon density over orientations weighted with
$|\Psi|^2$.

The density in the quantum state is therefore
\be
\rho_\Psi({\bf{x}})=\int
\mathcal{B}(D(A)^{-1}{\bf{x}})|\Psi(A)|^2\sin{\beta} \, d\alpha \,
d\beta \, d\gamma \,.
\label{19}
\ee
Here $A$ stands for the
$SU(2)$ matrix parametrized by Euler angles $\alpha$, $\beta$,
$\gamma$, and $D(A)$ for the $SO(3)$ matrix associated to $A$ via
\be
D(A)_{ab}=\frac{1}{2}{\rm Tr}(\tau_aA\tau_bA^\dag) \,.
\ee
As was already mentioned, the $B=2$ rational map is $R(z)=z^2$, and
this gives a good approximation to the $B=2$ Skyrmion solution. It
leads, using (\ref{6}), to the following expression for the
classical baryon density:
\be
\mathcal{B}(r,z)=\frac{1}{\pi}\left(\frac{1+|z|^2}{1+|z|^4}|z|\right)^2
g(r) \,,
\ee
where $g(r)$ is a radial function. $g(r)$ is
unaffected by the quantum averaging, so we ignore it from now on.
In terms of polar angles, the angular dependence of $\mathcal{B}$
is given by
\be
\mathcal{B}=\frac{1}{\pi}
\frac{(1+\tan^2{(\frac{\theta}{2}))}^2\tan^2{(\frac{\theta}{2})}}
{(1+\tan^4{(\frac{\theta}{2}}))^2} \,,
\ee
where this is normalized to
have angular integral equal to 2, the degree of the rational map.

To evaluate $\rho_\Psi({\bf{x}})$ we first expand $\mathcal{B}$ in
terms of spherical harmonics $Y_{lm}(\theta,\phi)$:
\be
\mathcal{B}=\sum_{l,m}{c_{lm}Y_{lm}(\theta,\phi)} \,, \label{7}
\ee
where, because of axial symmetry, there are only terms with
$m=0$. Although (\ref{7}) is an infinite series it is a good
approximation to take just the first two non--zero terms of the
sum,
\be
\mathcal{B}=c_{00}Y_{00}(\theta)+c_{20}Y_{20}(\theta) \,,
\label{classden}
\ee
as all the other terms contribute less than a $5\%$
correction. Because the map $R$ has degree 2, $c_{00} =
1/\sqrt{\pi}$; also, we find numerically that $c_{20} = -0.36$.
Then $\mathcal{B}(D(A)^{-1}\bf x)$ can be written as
\be
\mathcal{B}(\tilde{{\bf{x}}}) = c_{00}Y_{00}(\tilde{\theta})
+c_{20}Y_{20}(\tilde{\theta}) \,,
\ee
where $\tilde{\bf{x}}=D(A)^{-1}\,{\bf{x}}$ and similarly for
$\tilde{\theta}$, $\tilde{\phi}$. Using the transformation
properties of spherical harmonics under rotations,
\be
Y_{lm}(\tilde{\theta},
\tilde{\phi})=\sum_kD_{mk}^l(A)^*Y_{lk}(\theta,\phi) \,, \qquad
({\rm no  \,\, sum \,\, on \,\,}l) \,,
\ee
the fact that $|\Psi|^2
= (3/8\pi^2)D^1_{00}(A)D^1_{00}(A)^*$, the orthogonality
properties of the Wigner functions
\be
\int{D^j_{ab}(A)D^{j'}_{cd}}(A)^*\sin{\beta} \,  d\alpha \, d\beta
\, d\gamma=\frac{8\pi^2}{2j+1}\delta^{jj'}\delta_{ac}\delta_{bd} \,,
\ee
and (in terms of the Wigner $3j$ symbols)
\be
\int{D^j_{ab}(A)D^{j'}_{cd}(A)D^{j''}_{ef}(A)}\sin{\beta} \,
d\alpha \, d\beta \, d\gamma=8\pi^2\left(
\begin{array}{cccc}
j&j'&j''\\
a&c&e
\end{array}\right)\left(
\begin{array}{cccc}
j&j'&j''\\
b&d&f
\end{array} \right)\,,
\ee
we find directly from (\ref{19}) that the quantum probability
distribution is
\be
\rho_{\Psi}=c_{00}Y_{00}+\frac{2}{5}c_{20}Y_{20} \,.
\ee
This is an exact expression -- no higher terms are present. We see that it
resembles the classical distribution (\ref{classden}), but is more dominated
by the first term. Thus, when quantum effects are included, the classical
toroidal density remains, but is smoothed out to become more
spherically symmetric.

\section{$B=4$ case}

The minimal energy $B=4$ solution has cubic symmetry; the region
of high baryon density resembles a rounded cube with holes in the
faces and at the centre \cite{65}. We define the orthogonal
body--fixed axes to be those passing through the face centres, and
the standard orientation of the cube to be where these axes are
aligned with the Cartesian axes in space. We shall again consider the
Skyrmion as a rigid body, which means the configuration is not
allowed to vibrate. It was shown in \cite{411} that in this case
the ground state, representing the $\alpha$--particle, has quantum
numbers $i=0$ and $j=0$, with the (unnormalized) wavefunction
$\Psi^{(0)}=1$ being independent of the rotational and isospin
collective coordinates. The first excited state has $i=0$ and
$j=4$ and is \cite{Irwin}
\be
\Psi^{
(4)}_m=D^4_{4m}(\alpha,\beta,\gamma)
+\sqrt{\frac{14}{5}}D^4_{0m}(\alpha,\beta,\gamma)
+D^4_{-4m}(\alpha,\beta,\gamma) \,. \label{8}
\ee
In \cite{Irwin} the third component of the space--fixed spin, $m$,
was arbitrary. The structure of (\ref{8}) is required by the cubic
symmetry with respect to body--fixed axes.

But just fixing $m$ is not enough to make the wavefunction
cubically symmetric both on the left and on the right, i.e. also
with respect to space--fixed axes. To achieve this we need to take
the following linear combination of the above wavefunctions:
\be
\Psi^{(4)}=\Psi^{(4)}_4+\sqrt{\frac{14}{5}}\Psi^{(4)}_0+\Psi^{(4)}_{-4}
\,. \label{9}
\ee
The cubic symmetry in space is fairly
obvious by analogy with (\ref{8}), and can be verified as
follows. First note that symmetry under $90^\circ$ rotations about
the $x^3$-axis implies that all possible terms in (\ref{9}) with
$m$ other than $\pm4,0$ vanish. To simplify the calculations a bit
further we then introduce new variables
\be
a=\cos\Bigl({\frac{\beta}{2}}\Bigr)e^{\frac{1}{2}i\gamma}
e^{\frac{1}{2}i\alpha} \,, \quad
b=-\sin\Bigl({\frac{\beta}{2}}\Bigr)e^{-\frac{1}{2}i\gamma}
e^{\frac{1}{2}i\alpha} \,.
\ee
Obviously they satisfy
$|a|^2+|b|^2=1$. In terms of $a$ and $b$, the $SU(2)$ orientation
matrix parametrized by Euler angles $\alpha,\beta,\gamma$ is
$$
A=\left(
\begin{array}{cccc}
a&b\\
-\bar{b}&\bar{a}
\end{array}
\right) \,.
$$
In this notation the wavefunctions for different
$m$ take the following compact form:
\bea
\Psi^{(4)}_{4}&=&a^8+14a^4b^4+b^8 \nonumber \\
\Psi^{(4)}_0&=&\sqrt{70}\left(a^4\bar{b}^4+\bar{a}^4b^4
+ \frac{1}{40}(3-30(|a^2|-|b^2|)^2+35(|a^2|-|b^2|)^4)\right) \nonumber \\
\Psi^{(4)}_{-4}&=&\bar{a}^8+14\bar{a}^4\bar{b}^4+\bar{b}^8 \,.
\eea
Therefore the wavefunction (\ref{9}) in terms of $a$ and $b$ is
\bea
\Psi^{(4)}&=&2\textrm{Re}(a^8+14a^4b^4+b^8)+14(a^4\bar{b}^4+\bar{a}^4b^4)
\nonumber\\&&
+\,\,\frac{7}{20}\left(3-30(|a^2|-|b^2|)^2+35(|a^2|-|b^2|)^4\right)
\,.
\eea
As expected, it is real. By acting on $A$ with the
generators of the cubic group:
\begin{eqnarray}
\left(
\begin{array}{cccc}
 \frac{1+i}{\sqrt{2}}&0\\
0&\frac{1-i}{\sqrt{2}}
\end{array}
\right),\
\qquad \left(
 \begin{array}{cccc}
 \frac{1+i}{2}&\frac{1-i}{2}\\
-\frac{1+i}{2}&\frac{1-i}{2}
\end{array}
\right),
\end{eqnarray}
corresponding to a $90^{\circ}$ rotation around a face of the
cube, and a $120^{\circ}$ rotation around a diagonal of the
cube, we find the resulting transformations of $(a,b)$, and it is
easy to check that $\Psi^{(4)}$ is cubically symmetric both on the
left and on the right.

The wavefunction $\Psi^{(4)}$ has a positive maximum of $\frac{24}{5}$ at
the identity, $(a,b)=(1,0)$, and at all other elements of the
(double cover of the) cubic group. This is as desired, as it
corresponds to the Skyrmion having a high probability to be in its
standard orientation. But $\Psi^{(4)}$ also has a negative minimum
of $-\frac{104}{45}$, which gives a further local maximum of
$|\Psi^{(4)}|^2$, at an orientation obtained by a $60^\circ$
rotation around a diagonal of the cube, which is far from the
standard orientation. We wish to suppress this.

We can do this by being a bit more sophisticated than in the $B=2$
case. We still have the freedom of adding an arbitrary constant to
the wavefunction. This means taking a superposition of the ground
and first excited states, $\Psi^{(0)}$ and $\Psi^{(4)}$: \be \Psi
= \Psi^{(4)}+\kappa\Psi^{(0)} \,. \ee Here again we are interested
in the nucleon density of the configuration. Our goal will be to
adjust the constant $\kappa$ to get a quantum distribution as
close as possible to the classical one. As in the $B=2$ case we
define the quantum nuclear density via
\be
\rho_\Psi({\bf{x}})=\int
\mathcal{B}(D(A)^{-1}{\bf{x}})|\Psi(A)|^2\sin{\beta} \, d\alpha \,
d\beta \, d\gamma \,,
\ee
where $\mathcal{B}(\bf{x})$ is the
classical baryon density of the $B=4$ Skyrmion in its standard
orientation. Using the rational map (\ref{6^}), we find
that $\mathcal{B}$ has the angular dependence
\be
\mathcal{B}=\frac{12}{\pi}|z|^2(1+|z|^2)^2
\frac{(z^4\bar{z}^4-z^4-\bar{z}^4+1)}
{(z^4\bar{z}^4+z^4+12z^2\bar{z}^2+\bar{z}^4+1)^2} \,.
\ee
Expressed in terms of polar angles,
\bea
\mathcal{B}&=&\frac{12}{\pi}\tan^2\left(\frac{\theta}{2}\right)
\left(1+\tan^2\left(\frac{\theta}{2}\right)\right)^2\\\nonumber
&&\times
\,\,\frac{(\tan^8(\frac{\theta}{2})-2\tan^4(\frac{\theta}{2})\cos
4\phi +1)}{\left(\tan^8(\frac{\theta}{2}) +2\tan^4(\frac{\theta}{2})
\cos 4\phi +12\tan^4(\frac{\theta}{2})+1\right)^2} \,,
\eea
which may be expanded in the following form:
\be
\mathcal{B}=d_0Y_{00}+d_4Z_4(\theta,\phi)+d_6Z_6(\theta,\phi)+
d_8Z_8(\theta,\phi)+ \dots \,.
\ee
Here $Z_4, Z_6$ and $Z_8$ are the
unique cubically symmetric combinations of spherical harmonics
with, respectively $l=4,6$ and 8:\footnote{These can be derived
by combining the generating, cubically symmetric Cartesian
polynomials $x^2+y^2+z^2 \,$, $x^4+y^4+z^4 \,$, $x^6+y^6+z^6 \,$, and finding
the combinations which satisfy Laplace's equation \cite{n}.}
\bea
Z_4&=&Y_{44}+\sqrt{\frac{14}{5}}Y_{40}+Y_{4-4}\,, \nonumber \\
Z_6&=&Y_{64}-\sqrt{\frac{2}{7}}Y_{60}+Y_{6-4}\,, \nonumber \\
Z_8&=&Y_{88}+\sqrt{\frac{28}{65}}Y_{84}+
\sqrt{\frac{198}{65}}Y_{80}+\sqrt{\frac{28}{65}}Y_{8-4}+Y_{8-8} \,.
\eea
The leading
coefficient is $d_{0} = 2/\sqrt{\pi}$ because the rational map has
degree $4$, and by numerical calculation we find that $d_4=-0.28$,
$d_6=-0.032$ and $d_8=0.024$. Then, by a similar calculation as in
the $B=2$ case, normalizing the wavefunction and using the
orthogonality properties of the Wigner functions, we find the
following numerical result for the angular dependence of the
quantum baryon density:
\bea
\rho_\Psi=d_0Y_{00}+\frac{4}{2.56+\kappa^2} \Bigl\{-(0.038
+0.075\kappa)Z_4 -0.006Z_6+0.002Z_8\Bigr\}
  \,,
\label{b=4}
\eea
which is again a finite sum, all the further terms being zero.

In (\ref{b=4}), $\kappa$ is not yet specified. Let us adjust it in
such a way that the above distribution looks as close as possible
to the classical one, i.e. let us maximize the coefficient of the
$l=4$ terms:
$$\frac{4}{2.56+\kappa^2}(0.038+0.075\kappa) \,.$$
The maximum is at $\kappa\cong 1.17$, which leads to the following
expression for the quantum baryon density:
\bea
\rho_\Psi &\cong&
1.13Y_{00}-0.13Z_4-0.006Z_6+0.002Z_8 \nonumber \\
&\cong& d_0Y_{00}+0.46d_4Z_4+0.2d_6Z_6+0.1d_8Z_8 \,.
\eea

Thus in the $B=4$ case, as in the $B=2$ case, one can find a
quantum state which localizes the Skyrmion close to its standard
orientation, and which preserves the symmetry of the classical
solution. However, the inclusion of quantum effects smoothes the
classical baryon density, making it rather closer to spherically symmetric.
Again, the effect is to approximately halve the leading
non-constant harmonics, here with $l=4$. If we considered the pure
$j=4$ state $\Psi^{(4)}$, we would get
\be
\rho_{\Psi^{(4)}} \cong d_0Y_{00}+0.2d_4Z_4+0.3d_6Z_6+0.15d_8Z_8 \,,
\ee
which is much closer to spherically symmetric.

We can also find the energy of our state $\Psi$; it is
\be
E=\frac{1}{2.56+\kappa^2}\left(2.56 E_{j=4}+\kappa^2 E_{j=0}\right)
\cong 0.65E_{j=4}+0.35E_{j=0}\,,
\ee
so it is not as highly excited as a pure $j=4$ state.

The combination of $j=0$ and $j=4$ states, $\Psi$, is a bit
artificial as the quantum state of a free $B=4$ Skyrmion, but
would make sense if we were dealing with interacting Skyrmions
(for example, when describing compound nuclei such as $\rm{Be^8},
\rm{C^{12}}$ in the Skyrme model equivalent of the
$\alpha$--particle model). Here we expect the relative
orientations of the $B=4$ subclusters to be rather precisely
fixed when they are close together, so as to minimize their
potential energy.

\section{Conclusions}

Three well--localized wavefunctions of the $B=1$ Skyrmion have
been considered and some of their advantages and physical
implications have been discussed. The $B=2$ and $B=4$ minimal
energy Skyrmion solutions have been quantized in such a way that
the wavefunctions have the same symmetry properties as the
classical Skyrmions (respectively, axial and cubic symmetry both
on the left and on the right), and angularly localized quantum
states with shapes closest to the classical solutions have been
found. In the $B=4$ case, a superposition of two low--lying states
of definite angular momentum needed to be considered. It is
impossible to completely reproduce the shape of the classical
solution this way. The quantum state necessarily smoothes out the
classical baryon density, making it closer to being spherically
symmetric.

All our results were obtained in the rigid body approximation,
i.e. we did not allow the Skyrmions to vibrate. Considering the
vibrational modes may be an interesting topic for future work. The
vibrational modes for the $B=2$ and $B=4$ Skyrmions were
calculated in \cite{Turok,Barnes} and a qualitative analysis has
been given for $B=7$ \cite{Bask}.
The vibration frequencies obtained can be separated into those
below and those above the breather mode, which is the oscillation
corresponding to a change in scale size of the Skyrmion. In the
modes below the breather, the Skyrmion tends to splits up into
individual $B=1$ Skyrmions or small $B$ Skyrmion clusters.

But treating the vibration modes as harmonic oscillators is not
very accurate, since, as the minimal energy configuration
separates into individual Skyrmions the potential flattens out. A
more accurate treatment would involve estimating the
inter--Skyrmion potential at intermediate and large separations.
Thus it should not be expected that the inclusion of the zero
point energy of harmonic vibrational modes will yield accurate
results for masses, binding energies of states, etc.

The vibrational modes also couple to the rotational degrees of
freedom, which complicates the analysis of the rotational and
isospin wavefunctions \cite{vibr}.

\section*{Acknowledgements}
O.M. thanks EPSRC for the award of a Dorothy Hodgkin Scholarship.
N.S.M. thanks Nigel Cooper for a helpful discussion, and ECT*,
Trento for hospitality.

\end{document}